\documentclass{ws-p8-50x6-00}
\usepackage{graphicx}
\usepackage{amsmath}
\usepackage{amsfonts}
\usepackage{amssymb}

\begin{document}

\title{Modal field theory and quasi-sparse eigenvector diagonalization}
\author{Dean Lee}
\address{University of Massachusetts, Amherst, MA 01003  USA}
\maketitle
\abstracts{
We review recent developments in non-perturbative field theory using modal
field methods. \ We discuss Monte Carlo results as well as a new
diagonalization technique known as the quasi-sparse eigenvector method.}

\section{Introduction}

Of the many approaches to non-perturbative quantum field theory, we can
identify two general computational strategies. \ The first is the method of
Monte Carlo. \ The main advantages of this approach is that it can treat many
higher dimensional field theories, requires relatively little storage, and can
be performed with massively parallel computers. \ The other strategy is the
method of explicit diagonalization. \ The strong points of this method are
that it is immune to the fermion sign problem, can handle complex-valued
actions, and yields direct information about the spectrum and eigenstate wavefunctions.

Modal field theory is a simple Hamiltonian framework which can accomodate
either computational strategy. The first step is to approximate field theory
as a finite-dimensional quantum mechanical system. \ The approximation is
generated by decomposing field configurations into free wave modes and has
been explored using both spherical partial waves \cite{spher} and periodic box
modes \cite{periodic}. \ From there we can analyze the properties of the
reduced system using Monte Carlo, diagonalization, or some other computational
method. \ In this short review we present two different approaches using
the modal field formalism which address the example of $\phi^{4}$ theory in
$1+1$ dimensions. \ We begin with a summary of the Monte Carlo calculation of
the critical behavior of $\phi^{4}$ theory in Euclidean space. \ We then
discuss the computational challenges of a diagonalization-based approach and
conclude with a calculation of the lowest energy states of the theory using a
diagonalization technique known as the quasi-sparse eigenvector method.

\section{Monte Carlo}

In this section we review the basic features of modal field theory in a
periodic box and discuss spontaneous symmetry breaking in Euclidean
two-dimensional $\phi^{4}$ theory using the method of diffusion Monte Carlo.
\ The field configuration $\phi$ is subject to periodic boundary conditions
$\phi(t,x-L)=\phi(t,x+L).$ \ Expanding in terms of periodic box modes, we
have
\begin{equation}
\phi(t,x)=\sqrt{\tfrac{1}{2L}}%
{\displaystyle\sum\limits_{n=0,\pm1,...}}
\phi_{n}(t)e^{in\pi x/L}.
\end{equation}
The sum over momentum modes is regulated by choosing some large positive
number $N_{\max}$ and throwing out all high-momentum modes $\phi_{n}$ such
that $\left|  n\right|  >N_{\max}$. In this theory renormalization can be
implemented by normal ordering the $\phi^{4}$ interaction term. After a
straightforward calculation (details shown in \cite{periodic}), we find
that the counterterm Hamiltonian has the form
\begin{equation}
\tfrac{6\lambda b}{4!2L}%
{\displaystyle\sum\limits_{n=-N_{\max},... N_{\max}}}
\phi_{-n}\phi_{n},
\end{equation}
where
\begin{equation}
b=%
{\displaystyle\sum\limits_{n=-N_{\max},... N_{\max}}}
\tfrac{1}{2\omega_{n}},\qquad\omega_{n}=\sqrt{\tfrac{n^{2}\pi^{2}}{L^{2}}%
+\mu^{2}}.
\end{equation}
We represent the canonical conjugate pair $\phi_{n}$ and $\frac{d\phi_{-n}%
}{dt}$ using the Schr\"{o}dinger operators $q_{n}$ and $-i\tfrac{\partial
}{\partial q_{n}}$. \ Then the functional integral for $\phi^{4}$ theory is
equivalent to that for a quantum mechanical system with Hamiltonian
\begin{align}
H &  =%
{\displaystyle\sum\limits_{n=-N_{\max},... N_{\max}}}
\left[  -\tfrac{1}{2}\tfrac{\partial}{\partial q_{-n}}\tfrac{\partial
}{\partial q_{n}}+\tfrac{1}{2}(\omega_{n}^{2}-\tfrac{\lambda b}{4L}%
)q_{-n}q_{n}\right]  \label{hamilton}\\
&  +\tfrac{\lambda}{4!2L}%
{\displaystyle\sum\limits_{\genfrac{}{}{0pt}{1}{n_{i}=-N_{\max},... N_{\max
}}{n_{1}+n_{2}+n_{3}+n_{4}=0}}}
q_{n_{1}}q_{n_{2}}q_{n_{3}}q_{n_{4}}.\nonumber
\end{align}

The existence of a second-order phase transition in two-dimensional $\phi^{4}$
theory has been discussed in the literature \cite{glimm}. The phase
transition is due to $\phi$ developing a non-zero expectation value and the
spontaneous breaking of $\phi\rightarrow-\phi$ reflection symmetry. It is
believed that this theory belongs to the same universality class as the
two-dimensional Ising model and therefore shares the same critical exponents.
\ From the Ising model we expect to find
\begin{equation}
\nu=1,\quad\beta=\tfrac{1}{8}.\label{ising}%
\end{equation}

$\nu$ is the exponent associated with the inverse correlation length or,
equivalently, the mass of the one-particle state, $m$. \ It is defined by the
behavior of $m$ near the critical coupling $\lambda_{c},$%
\begin{equation}
m\sim(\lambda_{c}-\lambda)^{\nu}.
\end{equation}
We determine the behavior of the mass as we approach the critical point from
the symmetric phase of the theory. \ All computations are done using the
method of diffusion Monte Carlo (DMC). \ The idea of DMC is to model the
dynamics of the imaginary-time Schr\"{o}dinger equation using the diffusion
and decay/production of simulated particles. The kinetic energy term in the
Hamiltonian determines the diffusion rate of the simulated particles and the 
potential energy term determines the local decay/production
rate. A self-contained introduction to DMC can be found in \cite{kosztin}.

$\beta$ is the critical exponent describing the behavior of the vacuum
expectation value. \ In the symmetric phase the vacuum state is unique and
invariant under the reflection transformation $\phi\rightarrow-\phi$ (or
equivalently $q_{n}\rightarrow-q_{n},$ for each $n$)$.$ \ In the
broken-symmetry phase the vacuum is degenerate as $L\rightarrow\infty$,
and\ $q_{0}$, the zero-momentum mode, develops a vacuum expectation value.
\ In the $L\rightarrow\infty$ limit tunnelling between vacuum states is
forbidden. One ground state, $\left|  0^{+}\right\rangle ,$ is non-zero only
for values $q_{0}>0$ and the other, $\left|  0^{-}\right\rangle $, is non-zero
only for $q_{0}<0$. \ We will choose $\left|  0^{+}\right\rangle $ and
$\left|  0^{-}\right\rangle $ to be unit normalized. \ We can determine
$\beta$ from the behavior of the vacuum expectation value as we approach the
critical coupling,
\begin{equation}
\left\langle 0^{+}\right|  \phi\left|  0^{+}\right\rangle \sim(\lambda
-\lambda_{c})^{\beta}.
\end{equation}

In our calculations we used $(L,N_{\max})=(2.5\pi,$ $8),(2.5\pi,$ $10),(5\pi,$ $16),$ and $(5\pi,$ $20).$
For convenience we measure all quantities in units where $\mu=1$. 
For each set of parameters $L$ and $N_{\max},$ the curves for $m$ and
$\left\langle 0^{+}\right|  \phi\left|  0^{+}\right\rangle $ near the critical
coupling were fitted using the parameterized forms%

\begin{equation}
m=a\left(  \tfrac{\lambda_{c}^{m}}{4!}-\tfrac{\lambda}{4!}\right)  ^{\nu}%
\end{equation}
and
\begin{equation}
\left\langle 0^{+}\right|  \phi\left|  0^{+}\right\rangle =b\left(
\tfrac{\lambda}{4!}-\tfrac{\lambda_{c}^{\left\langle \phi\right\rangle }}%
{4!}\right)  ^{\beta}.
\end{equation}
Combining the results for the various values of $L$ and $N_{\max}$ and taking into account finite-size effects, we can extrapolate to the $L$, $N_{\max}\rightarrow\infty$ limit. \ The results we find are

\begin{align}
\tfrac{\lambda_{c}^{m}}{4!} &  =2.5\pm0.2\pm0.1,\ \nu=1.3\pm0.2\pm
0.1,\ a=0.43\pm0.05\pm0.02,\\
\tfrac{\lambda_{c}^{\left\langle \phi\right\rangle }}{4!} &  =2.5\pm
0.1\pm0.1,\ \beta=0.13\pm0.02\pm0.01,\ b=0.71\pm0.04\pm0.03.\nonumber
\end{align}
The first error bounds
include inaccuracies due to Monte Carlo statistics, higher energy states (for
the mass curves), and extrapolation. The second error bounds represent
estimates of the systematic errors due to our choice of initial state, time
step parameter, and bin sizes in the DMC simulations. \ Our results for the
critical exponents are consistent with the Ising model predictions
(\ref{ising}). \ The results for the critical coupling $\tfrac{\lambda_{c}%
^{m}}{4!}$ and $\tfrac{\lambda_{c}^{\left\langle \phi\right\rangle }}{4!}$ are
in agreement with a recent lattice computation \cite{loinaz}
\begin{equation}
\tfrac{\lambda_{c}}{4!}=2.56_{-.01}^{+.02}%
\end{equation}
as well as the discrete light-cone quantization result \cite{bender}%

\begin{equation}
\tfrac{\lambda_{c}}{4!}\approx\tfrac{(3+\sqrt{3})\pi}{6}\approx2.48.
\end{equation}

\section{Diagonalization}

In principle diagonalization techniques can provide significant detailed
information about a quantum system. \ The basic idea is to diagonalize the
Hamiltonian matrix defined over a truncated finite-dimensional Fock space.
\ In practise however few field theories are computationally feasible and
typically only theories in $1+1$ dimensions. \ Sparse matrix techniques such
as the Lanczos or Arnoldi schemes allow us to push the dimension of Fock space
to about 10$^{5}$ or 10$^{6}$ states. \ This may be sufficient to do accurate
calculations for $\phi_{1+1}^{4}$ theory near the critical point
$\frac{\lambda}{4!}\approx2.5$ for larger values of $L$ and $N_{\max}.$ \ It
is, however, near the upper limit of what is possible using current computer
technology and existing algorithms. \ Unfortunately field theories in $2+1$
and $3+1$ dimensions will require much larger Fock spaces, probably at least
10$^{12}$ and 10$^{18}$ states respectively. \ In order to tackle these larger
Fock spaces it is necessary to venture beyond standard diagonalization approaches.

In view of this difficulty one could ask\ how the Monte Carlo method avoids
this problem. \ The answer is by means of importance sampling.  While the space
of all configurations is impossibly large, it is often adequate to
sample only the most important configurations. \ Unfortunately Monte Carlo
techniques are not effective for a number of important problems. \ These
include systems with dynamical fermions and computations involving
complex-valued actions. \ In general difficulties arise when the functional
integral measure is not positive definite and produce contributions with
oscillating signs or phases.

In \cite{qse} a new importance sampling method was proposed for use within a
diagonalization scheme. \ The aim was to exploit the extreme sparsity of the
Hamiltonian matrix, a result of the restricted form that characterizes local
renormalizable interactions. \ If $N$ is the number of
non-zero entries per row or column in the Hamiltonian matrix$,$ it is shown in
\cite{qse} that a typical eigenvector is dominated by its largest
$\sqrt{N}$ elements.  There are some exceptions to this 
rule related to degeneracies which are discussed in
\cite{qse}. \ The algorithm for diagonalizing $H$ is as follows.\ \ We start
by choosing a set of orthonormal basis vectors such that the Hamiltonian
matrix $H_{ij}$ is sparse and the eigenvectors are \textit{quasi-sparse}.
\ The term \textit{quasi-sparse }indicates that the norm of a vector is
dominated by the contribution from a small fraction of its components. \ The
remaining steps are as follows:

\begin{enumerate}
\item  Select a subset of basis vectors $\left\{  e_{i_{1}},\cdots,e_{i_{n}%
}\right\}  $ and call the corresponding subspace $S$.

\item  Diagonalize $H$ restricted to $S$ and find one eigenvector $v.$

\item  Sort the basis components $\langle e_{i_{j}}|v\rangle$ according to
absolute size and throw away the least important basis vectors.

\item  Replace the discarded basis vectors by new basis vectors. \ These are
selected at random from a pool of candidate basis vectors which are connected
to the old basis vectors through non-vanishing matrix elements of $H$.

\item  Redefine $S$ as the subspace spanned by the updated set of basis
vectors and repeat steps 2 through 5.
\end{enumerate}

\noindent If the subset of basis vectors is sufficiently large, the exact
eigenvectors will be stable fixed points of the update process. \ We will
refer to this diagonalization technique as the quasi-sparse eigenvector (QSE) method.

We now test the QSE method on $\phi^{4}$ theory in $1+1$ dimensions. \ The
theory can be treated rather well using conventional sparse matrix techniques
and a Fock space of about 10$^{6}$ states. \ However the test we propose will
be more stringent. \ The size of Fock space and the number of non-zero
transitions per state for $\phi^{4}$ theory in $3+1$ dimensions are the cubes
of the corresponding numbers in $1+1$ dimensions. \ Therefore a more useful 
test is to solve the $1+1$ dimensional system in a manner that can be directly
generalized and expanded to the $3+1$ dimensional problem. \ Setting an upper
limit of $10^{6}$ states for the $3+1$ dimensional case, we will attempt to
solve the $1+1$ dimensional system using no more than 10$0$ (cube root of
10$^{6}$) basis vectors at a time.

We again use the Hamiltonian in (\ref{hamilton}) and set $L=5\pi$ and
$N_{\max}=20$. \ This corresponds with a momentum cutoff scale of
$\Lambda=4.$ \ We also place auxiliary constraints on the states in our
momentum Fock space. \ We keep only those states with $\leq13$ particles and
kinetic energy $\lesssim2\Lambda$. \ A precise definition of the kinetic
energy cutoff is provided in \cite{spectral}, and in terms of the parameter
$K_{\max}$ introduced there we use $K_{\max}=41$.\ In
our calculations we also restrict our attention to the zero momentum sector.

\begin{figure}[t]
\begin{center}
\epsfxsize=15pc 
\epsfbox{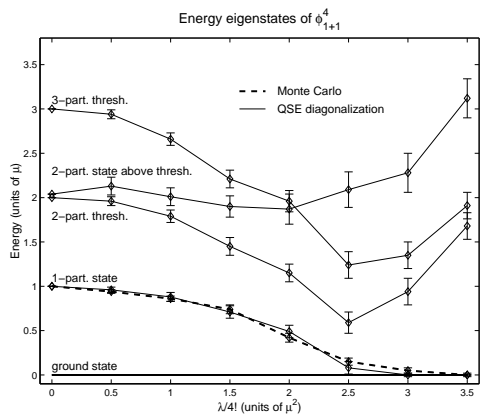} 
\caption{Energy eigenvalues of $\protect\phi _{1+1}^{4}$ as functions of the
coupling constant. \label{fig}}
\end{center}
\end{figure}

With these constraints our Fock space contains about $2\times10^{6}$ states
and the Hamiltonian matrix has about $10^{3}$ transitions per state.
\ However, we will restrict our QSE calculation to include only 100 basis
vectors at a time. \ Results for the energy eigenvalues are shown in
Fig.~\ref{fig}. \ From our Monte Carlo calculation we know that the
theory has a phase transition at $\frac{\lambda}{4!}\approx2.5$
corresponding with spontaneous breaking of the $\phi\rightarrow-\phi$
reflection symmetry. \ In the broken phase there are two degenerate ground
states and we refer to these as the even and odd vacuum states. \ In
Fig.~\ref{fig} we see signs of a second order phase transition near
$\frac{\lambda}{4!}\approx2.5$. \ Since we are working in a finite
volume the spectrum is discrete, and we can track the energy eigenvalues as
functions of the coupling. \ Crossing the phase boundary, we see that the
vacuum in the symmetric phase becomes the even vacuum in the broken phase
while the one-particle state in the symmetric phase becomes the odd vacuum.
\ The energy difference between the states is also in agreement with the Monte
Carlo calculation of the same quantities. \ The state marking the two-particle
threshold in the symmetric phase becomes the one-particle state above the odd
vacuum, while the state at the three-particle threshold becomes the
one-particle state above the even vacuum. \ These one-particle states should
be degenerate in the infinite volume limit. \ One rather unusual feature is
the behavior of the first two-particle state above threshold in the symmetric
phase. \ In the symmetric phase this state lies close to the two-particle
threshold. \ But as we cross the phase boundary the state which was the
two-particle threshold is changed into a one-particle state. \ Thus our
two-particle state is pushed up even further to become a two-particle state
above the even vacuum and we see a pronounced level crossing.

A conventional sparse matrix algorithm would require about $10$ GB RAM and
10$^{5}$ Gflops to determine the energy and wavefunctions of the lowest lying
states for this system. \ The QSE method was able to perform the same task
with about $30$ KB RAM and 10$^{2}$ Gflops. \ The reductions in floating point
operations and memory are about three and six orders of magnitude respectively
and show the advantages of QSE diagonalization over other numerical
techniques. \ Other large non-perturbative systems should now also be amenable
to numerical solution using this method.

\section*{Acknowledgments}
I thank my collaborators on the works cited here and the
organizers and participants of the Workshop on Light-Cone QCD and
Nonperturbative Hadron Physics in Adelaide. \ Financial support provided by
the National Science Foundation.

\end{document}